\documentstyle[12pt]{article}
\begin{document}

\begin{flushright}
PSI-PR-94-34
nucl-th/9410043
\end{flushright}

\begin{center}
{\bf
\large
Isospin Recoupling and
Bose-Einstein Pion Correlations \\
in $\bar{N}N$ Annihilations }
\end{center}
\normalsize
\medskip

\begin{center}
R. D. Amado,$^{a,b}$ F. Cannata,$^{b,c}$ J-P. Dedonder,$^{b,d}$
M. P. Locher,$^{b}$ \\ and Yang Lu$^{a,b}$\\
\end{center}
$^a$ {\it Department of Physics, University of Pennsylvania,\\
 Philadelphia, PA 19104, USA      \\}
$^b$ {\it Paul Scherrer Institute, CH-5232 Villigen PSI, Switzerland\\}
$^c$ {\it Dipartimento di Fisica and INFN \\
  I-40126 Bologna, Italy \\}
$^d$ {\it Laboratoire de Physique Nucl\'{e}aire, Universit\'{e}
Paris 7-Denis Diderot, \\
   2 Place Jussieu, F-75251 Paris Cedex 05 \\
and Division de Physique Th\'{e}orique, IPN \\
F-91406 Orsay, France \footnote{The Division de
Physique Th\'{e}orique is a Research Unit of the
Universities of Paris 11 and 6 associated to CNRS.
}}\\

\centerline{{\bf Abstract}}
\noindent
We study the effect of isospin projections on a two source picture of
coherent pions coming from nucleon-antinucleon annihilation. We show
that important Hanbury-Brown Twiss correlations among like charge pion
pairs compared with unlike arise from these isospin projections.
\noindent
\newpage

In a recent letter \cite{ACDLL}, we showed that intensity correlations
among pions from $\bar{N}$$N$ annihilation at rest
can show strong Hanbury-Brown
Twiss \cite{H-BT} (H-BT)
correlations for like charged pion pairs without
having to invoke any thermal or related random phase assumptions.
Integration over the pion sources, experimental binning and isospin
projections are enough to produce a relative enhancement between like
pions pairs at small relative momentum compared with unlike pairs.
In that discussion we treated isospin rather cursorily, averaging
the  isospin of the pion field independently at each field emission
point.  We reported there, without proof, that if more careful isospin
projections are performed, the central results do not change.
The purpose of this note is to present the calculation that
substantiates that claim.
We will assume two sources of pion radiation, each emitting an ensemble
of pions restricted to a fixed isospin. Coupling the two isospins to
a fixed total isospin we shall show that there is an
enhancement at small relative
momentum of two pions with like charge.

Our presentation will be in terms of coherent states and will
use the isospin projection formalism we
exploited before \cite{BSS,ACDLS}.  We will also invoke
results from our letter \cite{ACDLL}.  Hence this note will not
stand by itself and the preceding papers should be consulted.

Consider the coherent state for pion radiation from two sources.  The
sources each have the same Fourier profile, $f(\vec{p})$, but
are displaced in space and perhaps time from  each other.
One source emits pions
in a field of
isospin $I_1$ and the other in isospin $I_2$.  These two are coupled
up to the total isospin $I$ and z-component $M$ of the entire system.
That state can be written \cite{ACDLS}

\begin{eqnarray}
|f,I_1,I_2;I,M>&=&
\sum_{M_{1},M_{2}}
<IM|I_1 M_{1}; I_2 M_{2}>
\int d\hat{T}_1 d\hat{T}_2\;
Y_{I_1 M_{1}}^{\ast}(\hat{T}_1)
Y_{I_2 M_{2}}^{\ast}(\hat{T}_2)\nonumber \\
&&\exp\left[\int d^3p f(\vec{p}) \exp (i \phi_1(p)) \hat{T}_1
\cdot \vec{a}^{\dagger}(\vec{p}\,)\right]
\nonumber \\
& &
\exp\left[\int d^3p f(\vec{p}) \exp (i \phi_2(p)) \hat{T}_2
\cdot \vec{a}^{\dagger}(\vec{p}\,)\right] |0>
\end{eqnarray}
where the phases $\phi_i(p)$ ($i=1,2$) are
\begin{equation}
\phi_i (p)=\tilde{\phi_i}+ \omega_p t_i-\vec{p}\cdot\vec{x}_i.
\end{equation}
and where $\vec{a}^{\dagger}(\vec{p})$ is the isovector creation
operator of a pion of momentum $\vec{p}$.  Here
$\tilde{\phi_i}$ is a phase associated with site $i$. It is
introduced to represent the effect of the various averaging mechanisms
discussed in \cite{ACDLL}.
As we explain there,  these averages have the same effect as random
phases of thermal or other origin,  but arise in the
purely coherent sources we consider.  Rather than repeating the
discussion from \cite{ACDLL} here, we introduce the
phases $\tilde{\phi_i}$ with the property
$<e^{i\tilde{\phi_i}} e^{-i\tilde{\phi_j}}>=\delta_{ij}$.
In all matrix elements we write, we understand this average to
take place. This average replaces all the summing, binning, etc.
we discussed in \cite{ACDLL}.
It should be noted that the state we have defined in (1)
is not normalized, but that the normalization
will cancel when we take ratios.

Consider first the one pion spectrum. To save writing let
us call the state defined in (1) $|f>$.  The density of pions of
charge type $\mu$  and momentum $\vec{k}$ in that state is given by
\begin{eqnarray}
W_1(\vec{k},\mu)&=&
<f|a_{\mu}^{\dagger}(\vec{k})a_{\mu}(\vec{k})|f>
\nonumber \\
&=&|f(\vec{k})|^2
\sum_{(1)(2)}
<IM|I_1 M_{1}; I_2 M_{2}>
<I'M'|I_1' M_{1}'; I_2' M_{2}'>
\nonumber \\
& & \int
d \hat{T}_1
d \hat{T}_2
d \hat{T}_1'
d \hat{T}_2'\;\;
Y_{I_1 M_{1}}^{\ast} (\hat{T}_1)
Y_{I_2 M_{2}}^{\ast} (\hat{T}_2)
Y_{I_1' M_{1}'}(\hat{T}_1')
Y_{I_2' M_{2}'}(\hat{T}_2')
\nonumber \\
& & \exp\left[N_0 ( \hat{T}_1\cdot \hat{T}_1'
+ \hat{T}_2\cdot \hat{T}_2')\right]
\left[T_{1\mu}^{\ast}T_{1\mu}'+T_{2\mu}^{\ast}T_{2\mu}'\right]
\end{eqnarray}
where $\sum_{(1)(2)}$ is over indices
$M_{1},M_{2}, M_{1}'$ and $M_{2}'$. The mean
number of pions emitted by each source,
$N_0$, is given by $\int d^3k |f(\vec{k})|^2$.
In obtaining this expression, we have used the phase
averaging discussed above both in the matrix element and
in the exponent.

The corresponding expression for the two pion spectrum,
that is for the probability of finding a pion of
momentum $\vec{p}$ and charge type $\mu$ and another of
momentum $\vec{q}$ and charge type $\nu$
is given by
\begin{eqnarray}
W_2(\vec{p},\,\mu;\vec{q},\nu)&=&
<f| a_{\mu}^{\dagger}(p)a_{\mu}(p)
a_{\nu}^{\dagger}(q)a_{\nu}(q) |f>
\nonumber \\
&=&
|f(\vec{p})|^2 |f(\vec{q})|^2
\sum_{(1)(2)}
<IM|I_1 M_{1}; I_2 M_{2}>
<I'M'|I_1' M_{1}'; I_2' M_{2}'>
\nonumber \\
& & \int
d \hat{T}_1
d \hat{T}_2
d \hat{T}_1'
d \hat{T}_2'\;\;
Y_{I_1 M_{1}}^{\ast} (\hat{T}_1)
Y_{I_2 M_{2}}^{\ast} (\hat{T}_2)
Y_{I_1' M_{1}'}(\hat{T}_1')
Y_{I_2' M_{2}'}(\hat{T}_2')
\nonumber \\
& & \exp\left[N_0 ( \hat{T}_1\cdot \hat{T}_1'
+ \hat{T}_2\cdot \hat{T}_2')\right]
\nonumber \\ & &
\left[
T_{1\mu}^{\ast} T_{1\nu}^{\ast} T_{1\mu}' T_{1\nu}'
+(1\leftrightarrow 2)
\right.
\nonumber \\ & &
+T_{1\mu}^{\ast} T_{2\nu}^{\ast} T_{1\mu}' T_{2\nu}'
+(1\leftrightarrow 2)
\nonumber \\ & &
\left.
+T_{1\mu}^{\ast} T_{2\nu}^{\ast} T_{2\mu}' T_{1\nu}'
{\rm e}^{-i\Delta E(t_1-t_2)+i\vec{Q}\cdot(\vec{x}_1-\vec{x}_2)}
+(1\leftrightarrow 2)
\right]
\end{eqnarray}
where $\Delta E=\omega_p-\omega_q$ and $\vec{Q}=\vec{p}-\vec{q}$
and where again $\tilde{\phi}$ averaging has been used.

The integrals in the one and two pion densities can be done, but
they are complicated.  They can be greatly simplified by taking
advantage of the fact that the mean number of pions, $N_0$ is large.
If we include the normalization, we have the factor
\begin{equation}
\exp\left(N_0(\hat{T}_1^{\ast}\cdot\hat{T}_2-1)\right)
\end{equation}
in our integrals.  For large $N_0$ this is well approximated by
\begin{equation}
\delta(
\hat{T}_1-\hat{T}_2)
\end{equation}
with corrections of order $1/N_0$.
In this limit,
only two integrals appear in our calculation of
the pion spectra,
\begin{equation}
A^{I_1 M_{1};I_2 M_{2}}_{\mu\nu}=
\int d\hat{T}\;
Y_{I_1 M_{1}}^{\ast}(\hat{T})
Y_{I_2 M_{2}}(\hat{T})T_{\mu}^{\ast}T_{\nu}
\end{equation}
and
\begin{equation}
B^{I_1 M_{1};I_2 M_{2}}_{\mu\nu}=
\int d\hat{T}\;
Y_{I_1 M_{1}}^{\ast}(\hat{T})
Y_{I_2 M_{2}}(\hat{T})
T_{\mu}^{\ast}
T_{\nu}^{\ast}T_{\mu}
T_{\nu}.
\end{equation}
A simple calculation yields
\begin{eqnarray}
A^{I_1 M_{1};I_2 M_{2}}_{\mu\nu}&=&
\sum_{LM}\frac{\sqrt{(2I_1+1)(2 I_2+1)}}{2L+1}
\nonumber \\
&&
<L0|I_1 0;10>
<L0|I_2 0;10>
\nonumber \\
&&
<LM|I_1 M_{1};1\mu>
<LM|I_2 M_{2};1\nu>
\end{eqnarray}
and
\begin{eqnarray}
B^{I_1 M_{1};I_2 M_{2}}_{\mu\nu}&=&
\sum_{LMa\alpha b\beta }\frac{\sqrt{(2I_1+1)(2 I_2+1)}}{2L+1}
\nonumber \\
&&
<a0|1 0;10>
<b0|1 0;10>
\nonumber \\
&&
<a\alpha|1 \mu;1\nu>
<b\beta|1 \mu;1\nu>
\nonumber \\
&&
<L0|a 0;I_1 0>
<L0|b 0;I_2 0>
\nonumber \\
&&
<LM|a \alpha;I_1 M_{1}>
<LM|b \beta;I_2 M_{2}>.
\end{eqnarray}
Note that in $A$ and $B$,  $I_1$ and $I_2$  must
have the same ``parity," that is they must be both even or
both odd.

In terms of the quantities $A$ and $B$, we can write

\begin{eqnarray}
W_1(\vec{k},\mu)&=&|f(k)|^2\sum_{(1)(2)}
<IM|I_1 M_{1}; I_2 M_{2}>
<I'M'|I_1' M_{1}'; I_2' M_{2}'>
\nonumber \\
&&
\left[
A^{I_1 M_{1};I_{1}'M_{1}'}_{\mu\mu}
\delta_{I_2I_2'}\delta_{M_{2}M_{2}'}
+(1\leftrightarrow 2)
\right]
\end{eqnarray}
and
\begin{eqnarray}
W_2(\vec{p},\mu;\vec{q}, \nu)&=& |f(p)|^2 |f(q)|^2
\sum_{(1)(2)}
<IM|I_1 M_{1}; I_2 M_{2}>
<I'M'|I_1' M_{1}'; I_2' M_{2}'>
\nonumber \\
&&
\left[
B^{I_1 M_{1};I_{1}'M_{1}'}_{\mu\nu}
\delta_{I_2I_2'}\delta_{M_{2}M_{2}'}
+(1\leftrightarrow 2)
\right.
\nonumber \\
&&
A^{I_1 M_{1};I_{1}'M_{1}'}_{\mu\mu}
A^{I_2 M_{2};I_{2}'M_{2}'}_{\nu\nu}
+(1\leftrightarrow 2)
\nonumber \\
&&
\left.
A^{I_1 M_{1};I_{1}'M_{1}'}_{\mu\nu}
A^{I_2 M_{2};I_{2}'M_{2}'}_{\nu\mu}
{\rm e}^{-i\Delta E(t_1-t_2)
+i\vec{Q}\cdot(\vec{x}_1-\vec{x}_2)}
+(1\leftrightarrow 2)
\right].
\end{eqnarray}

We now explore the consequences of these formulae for pion
correlations.  To simplify matters and to conform to our
physical prejudices, we limit all isospins to be less than
or equal to 1.  That is we assume there are no isotensor or
higher waves.
We are interested in the two pion correlations, $W_2$\footnote{
The one-pion spectrum (11) can also be calculated. In the case of
$I=I'=0$, we have $W_1(+)=W_1(-)=W_1(0)=2/3\;|f(k)|^2$.}.
For diagonal elements ($I=I'$), we can write\footnote{
Actually, it is $W_2/(|f(p)|^2|f(q)|^2)$ which is shown in
(13). However, the overall factor cancels when ratios are taken,
see (17).}

\begin{equation}
W_2=C+D\;
\cos[\Delta E(t_1-t_2)-\vec{Q}\cdot(\vec{x}_1-\vec{x}_2)].
\end{equation}
For Hanbury-Brown Twiss correlations, it is the ratio
$D/C$ that counts.  In Table 1 we show the values of
$C$ and $D$ for various choices of $I,M,I_1$, and $I_2$ for
different charges of the two pions. Since both $C$ and $D$
have the property $(-\mu,-\nu)=(\nu,\mu)=(\mu,\nu)$,
we show only one case for each $\mu\nu$.
We see that in many cases, $D=0$ for unlike charges
while it is not $0$ for like charges,
leading to H-BT correlations between like pairs.
In those cases in which $D$ is not zero, the ratio $D/C$
is considerably larger for the like charge case than for the unlike
(there are even some unlike cases where the ratio is negative)
again leading to important H-BT correlations. Thus we see
that, as asserted in \cite{ACDLL}, the effect of isospin projections
is to favor H-BT correlations among like pion pairs.

For annihilation from the $\overline{p} p$ system, the initial
state is not a pure isospin state but rather a superposition
of the $I=0,M=0$ and the $I=1,M=0$ states. In the absence of
initial state interactions this
state is $|\overline{p}p>=1/\sqrt{2}(|I=1,M=0>-
|I=0,M=0>)$. The  two-pion correlation function
from this $\overline{p}p$ initial state under the 2-source assumption
is then, schematically
\begin{equation}
W_2(\overline{p}p)=\frac{1}{2}(<1|W|1>+<0|W|0>-<1|W|0>-<0|W|1>)
\end{equation}
This involves off-diagonal terms in $I$ and $I'$.
Their contribution to
$W_2$ may be written
\begin{equation}
W_2=C+i\;D\;
\sin[\Delta E(t_1-t_2)-\vec{Q}\cdot(\vec{x}_1-\vec{x}_2)]
\end{equation}
The relevant $C$ and $D$ are shown in Table 2.
Note that $C=0$ and
the coefficients $D$ have the symmetry:
\begin{equation}
D(\mu,\nu)=-D(\nu,\mu)=-D(-\mu,-\nu)
\end{equation}

One observes that the symmetry properties of the off-diagonal
elements are such that they cancel each other in (14). However, they
will not cancel if the
initial state distortions make the relative phase between the
isospin 0 and 1 amplitudes complex.

To get a sense of the size of the H-BT correlations let us write
the ratio of like to unlike charge pions two particle pion spectra
as
\begin{equation}
\frac{W_2(\vec{p},+;\vec{q},+)}{W_2(\vec{p},+;\vec{q},-)} =
1+H \; \cos[\Delta E(t_1-t_2)-\vec{Q}\cdot(\vec{x}_1-\vec{x}_2)]
\end{equation}
where the size of $H$ controls the size of the H-BT correlations.
We can calculate $H$ for some typical situations from Table 1.
(For those cases in which $D$ is not zero  for the unlike charges,
it is always small compared with $C$ and we expand up from the
denominator to obtain $H$.)  For $I=I'$ the values of $H$ are
given in Table 3.  We see that for all cases, $H$ is of reasonable
size and for some cases it is quite large. Thus the isospin
projections lead to important H-BT correlations among the pions.

In summary,
we have seen that for two sources  the effects of isospin projections
when combined with the other averaging mechanisms we discussed in
\cite{ACDLL} give strong H-BT enhancements among like
charge pions from annihilation compared with unlike charge.
For one of the cases we discuss here ($I=0$ with $I_1=I_2=0$) the
argument is the same as the isospin averaging of \cite{ACDLL}.
If one sums over all isospins and uses completeness, the result
for $W_2$ again reduces to the isospin averaging of \cite{ACDLL}.
That it holds for individual isospins
is a happy but not obvious fact.

RDA, FC, and J-PD thank the theory group of the Division of Nuclear
and Particle Physics of the Paul Scherrer Institute for, once again,
providing a stimulating environment for research.
The work of RDA is partially
supported by the United States National Science Foundation.

\begin{table}
\begin{center}
\caption{Diagonal matrix element for two pion spectrum.}
\vspace*{0.5cm}
\begin{tabular}{|cccccc|c c|}\hline
$I$     &     $M$    &    $I_1$    &    $I_2$   &   $\mu$  &
$\nu$   &   C   & D
\\ \hline
 0 & 0 & 0 & 0 &+&+& 22/45 & 2/9 \\
   &   &   &   &+&--& 22/45 & 0 \\
   &   &   &   &+&0& 16/45 & 0 \\
   &   &   &   &0&0& 28/45 & 2/9 \\ \hline
 1 & 0 & 1 & 0 &+&+& 34/105 & 2/15 \\
   &   &   &   &+&--& 34/105 & 0 \\
   &   &   &   &+&0& 44/105 & 0 \\
   &   &   &   &0&0& 36/35 & 2/5 \\ \hline
 1 & 1 & 1 & 0 &+&+& 4/7 & 4/15 \\
   &   &   &   &+&--& 4/7 & 0 \\
   &   &   &   &+&0& 34/105 & 0 \\
   &   &   &   &0&0& 44/105 & 2/15 \\ \hline
 0 & 0 & 1 & 1 &+&+& 38/75 & 6/25 \\
   &   &   &   &+&--& 38/75 & 8/75 \\
   &   &   &   &+&0& 8/25 & 2/75 \\
   &   &   &   &0&0& 52/75 & 22/75 \\ \hline
 1 & 0 & 1 & 1 &+&+& 116/175 & 8/25 \\
   &   &   &   &+&--& 116/175 & --4/25 \\
   &   &   &   &+&0& 48/175 & 0 \\
   &   &   &   &0&0& 44/175 & 2/25 \\ \hline
 1 & 1 & 1 & 1 &+&+& 68/175 & 4/25 \\
   &   &   &   &+&--& 68/175 & 0 \\
   &   &   &   &+&0& 74/175 & --1/25 \\
   &   &   &   &0&0& 132/175 & 6/25 \\ \hline
\end{tabular}
\end{center}
\end{table}

\begin{table}
\caption{Off-diagonal matrix elements of two pion spectrum.}
\begin{center}
\vspace*{0.5cm}
\begin{tabular}{|cccccccc|c c|}\hline
$I$     &     $M$    &  $I'$ & $M'$ &  $I_1$    &    $I_2$   &
$\mu$   &     $\nu$  &  $C$  & $D$
\\ \hline
 1 & 0 & 0 & 0 & 1 & 1 &+&--& 0 & $4\sqrt{6}/75$\\
   &   &   &   &   &   &+&0& 0 & $\sqrt{6}/75$\\ \hline
\end{tabular}
\end{center}
\end{table}

\begin{table}
\caption{Coefficient $H$ of the Bose-Einstein term calculated from
Table 1.}
\begin{center}
\vspace*{0.5cm}
\begin{tabular}{|cccc|c|}\hline
$I$     &     $M$    &  $I_1$    &    $I_2$   &   $H$
\\ \hline
 0 & 0 & 0 & 0 & $5/11$ \\
 1 & 0 & 1 & 0 & $7/17$ \\
 1 & 1 & 1 & 0 & $7/15$ \\
 0 & 0 & 1 & 1 & $5/19$ \\
 1 & 0 & 1 & 1 & $21/29$ \\
 1 & 1 & 1 & 1 & $7/17$ \\ \hline
\end{tabular}
\end{center}
\end{table}

\begin{thebibliography}{hg}
\bibitem{ACDLL} R.D. Amado, F. Cannata, J-P. Dedonder, M.P. Locher
and Y. Lu, PSI-PR-94-27, to appear in Phys.Lett.{\bf B}.
\bibitem{H-BT} R. Hanbury-Brown and R. Q. Twiss, Philos. Mag.
{\bf 45} 633 (1954).
\bibitem{BSS} J.C. Botke, D.J. Scalapino, and R.L. Sugar,
Phys.Rev {\bf D9}, 813 (1974).
\bibitem{ACDLS} R.D. Amado, F. Cannata, J-P. Dedonder, M.P. Locher,
and B. Shao,
Phys. Rev. {\bf C50}, 640 (1994).
\end{thebibliography}
\end{document}